\documentclass[aps,prl,amsfonts,twocolumn,superscriptaddress]{revtex4}
\usepackage{amssymb}
\usepackage{epsfig}
\usepackage[cp1250]{inputenc}
\begin{document}
\newtheorem{proposition}{Proposition}
\newtheorem{lemma}{Lemma}
\newtheorem{fact}{Fact}
\newtheorem{definition}{Definition}
\newtheorem{example}{Example}

\title{Package of facts and theorems for efficiently generating entanglement criteria for many qubits}
\author{Marcin Wie\'sniak}\affiliation{Institute of Theoretical Physics and Astrophysics, University of Gda\'nsk,\\ ul. Wita Stwosza 57, PL-80-952 Gda\'nsk, Poland}
\author{Koji Maruyama}\affiliation{Department of Chemistry and Materials Science, Osaka City University,\\ Sugimoto, Osaka 558-8585  Japan}

\begin{abstract}
We present a package of mathematical theorems, which allows us to construct multipartite entanglement criteria. Importantly, establishing bounds for certain classes of entanglement does not take an optimization over continuous sets of states. These bonds are found from the properties of commutativity graphs of operators used in the criterion. We present two examples of criteria constructed according to our method. One of them detects genuine 5-qubit entanglement without ever referring to correlations between all five qubits.
\end{abstract}
\maketitle
For pure states, the notion of entanglement is relatively simple to define, even for multipartite cases. A composite system is said to be entangled if subsystems are in certain states only in reference to each other. For genuine multipartite entanglement we need to define properties of all but one subsystems via projective measurements to have the last one in a pure state. In such situations, all reduced density matrices are mixed.

The problem becomes more involved when we extend our consideration to mixed states. We can no longer rely on the contrast in purities of the global and reduced states. Instead, we need to refer to the operational definition. An $N$-particle mixed state is genuinely multipartite entangled if we cannot obtain it by local actions, classical communications, and distributing states that are separable with respect to any division of subsystems.

The difficulties we face in discussing multipartite entanglement have more than one origin.  One problem is the relatively large complexity of the question. If we want to claim that a state is multipartite-entangled, not only do we need to demonstrate that it is not separable with respect to any cut, but we also need to show that a state is not a mixture of separable states with different cuts. The other problem is that multipartite entanglement can come in various classes of equivalence. For example, D\"ur {\em et al.} \cite{DUR} have demonstrated that a collection of 3-qubit W states cannot be transformed by local operations and classical communication into a collection of 3-qubit GreenBerger-Horne-Zeilinger (GHZ) states, or vice versa.
The difference between such classes of multipartite entanglement is yet to be fully understood. The first step toward such an understanding is to deliver a quantifier of entanglement vanishing for one such class and not for some other \cite{OSTERLOH}.

Nevertheless, a number of multipartite-entanglement criteria have been developed so far. To give a few examples, one of the first criteria is based on the observation that states with some separability cannot violate Werner-Wolf-\.Zukowski-Brukner (WWW\.ZB) inequalities \cite{WW,WZ,ZB} as strongly as genuinely multipartite-entangled states \cite{YuChen}. More conventional methods, such as entanglement witnesses \cite{WIT1,WIT2}, have also been employed experimentally \cite{WIT3}.  For systems with continuous variables, various techniques to distinguish multipartite-entanglement have been proposed \cite{CV}. Entanglement quantifiers, such as concurrence, have been generalized to multipartite cases as well \cite{KUS}. Laskowski and \.Zukowski gave a simple criterion to distinguish genuine $N$-particle quantum correlations and those involving fewer particles \cite{LZ}, using generalized Bell inequalities based on a single entry of a density matrix. More recently, nonlinear multipartite entanglement criteria were presented in, e.g., \cite{EXPFRIEN,EXPFRIEN1}, in which the arguments were based on a simple geometrical fact concerning the correlation functions that can be determined by local experiments.

Here we present a useful package of mathematical facts and theorems, which allows us to analyze entanglement criteria for many qubits in terms of the squares of correlation tensor elements. We will give an example of its applications to construct a 5-partite entanglement criterion exploiting only four-partite correlations. In this way, we can demonstrate ``glocal'' aspects of some states detected by this criterion. On the one hand, entanglement is {\em globally} distributed over all qubits. On the other hand, it manifests in correlations between not all subsystems (hence {\em local}). Moreover, the relation between multipartite entanglement (ME) and correlations of a given order is very loose. For example, all reduced states of generalized GHZ states \cite{GHZ} are purely classical; we cannot conclude the presence of ME from them without a promise of the purity of the global state. The Smolin State \cite{SMOLIN}, on the other hand, has no lower-order correlations, while it is not multipartite-entangled. In this respect, genuinely multipartite-entangled states without $N$-qubit correlations have been illustrated in \cite{SENS}.

Before we start filling out our toolbox, we need to formulate our criterion. We discuss the states of $N$ qubits. A convenient representation of such states is through the correlation tensor, $\rho=\frac{1}{2^N}\left(\sum_{a,b,c,...=0}^3T_{abc...}\sigma_a\otimes\sigma_b\otimes\sigma_c\otimes...\right)$, where $\sigma_0$ is a $2\times 2$ unit matrix (later occasionally denoted simply as 1), $\sigma_1\equiv x\equiv\left(\begin{array}{cc}0&1\\1&0\end{array}\right),\sigma_2\equiv y\equiv\left(\begin{array}{cc}0&-i\\i&0\end{array}\right),\sigma_3\equiv z\equiv\left(\begin{array}{cc}1&0\\0&-1\end{array}\right)$. Tensor elements can be obtained from an experiment, $T_{abc...}=\text{tr}\rho(\sigma_a\otimes\sigma_b\otimes\sigma_c...)$. Out of the set of $4^N$ possible operators $\sigma_a\otimes\sigma_b\otimes\sigma_c...$ we choose some subset $\sigma$ and let $s$ denote its elements. [Hereafter, the tensor product sign ($\otimes$) will be omitted.]
The separability criteria are written in the form of an inequality:
\begin{equation}
\label{bellcrit}
Q=\sum_{s\in\sigma}(\text{tr}(\rho s))^2>B_{SC}.
\end{equation}
When Eq. (\ref{bellcrit}) holds, the state cannot belong to the separability class $SC$. The bound $B_{SC}$ is dependent on the considered $SC$ and $\sigma$. The question is, how can we find $B_{SC}$ in an efficient way?

Let us now present our package, which will allow us to conveniently obtain $B_{SC}$'s. It contains two definitions, three facts, and four lemmas.

\begin{fact}

Two tensor products of element of the Pauli group $\left\{1,x,y,z\right\}$, either commute or anticommute.
\end{fact}
They anticommute whenever they differ at odd number of positions, excluding pairs involving the unit operator. For example, $xxx$ commutes with $xyz$ and $yz1$, while it anticommutes with $x1y$.

\begin{lemma}

Let $\vec{O}=(O_1,...,O_n)$ be a vector of traceless Hermitian operators with eigenvalues $\pm 1$, which all anticommute with each other. Then for any (quantum mechanical) state we have $\sum_{i=1}^N\langle O_i\rangle^2\leq 1$.
\end{lemma}

{\em Proof:} Despite a proof presented in Ref. \cite{mono}, we provide an independent argument here. First we show that $\vec{V}\cdot\vec{O}$ has eigenvalues $\pm 1$, provided that $\vec{V}$ is a real vector of the unit length. Consider
\begin{equation}
(\vec{V}\cdot\vec{O})^2=\vec{V}^2I=I,
\end{equation}
where $I$ stands for the unit matrix of a respective dimension, $d$. All the cross terms in the square of $\vec{V}\cdot\vec{O}$ vanish by anticommutativity, and the operator $\vec{V}\cdot\vec{O}$ is traceless by the tracelessness of $O_i$'s. Any quantum mechanical states of this dimension can be written as
\begin{equation}
\rho=\frac{1}{d}\left(I+\vec{r}\cdot\vec{O}+\vec{s}\cdot\vec{P}\right).
\end{equation}
$\vec{P}=(P_1,...,P_M)$ is a vector of hermitian traceless operators, which complete an orthogonal basis of operators. We demand that 
$\text{tr}(P_j)=\text{tr}(O_iP_j)=\text{tr}(P_jP_k)=0$ for $j\neq k$.
The mean value of $\vec{V}\cdot\vec{O}$ with respect to a state $\rho$ is  $\text{tr}\rho(\vec{V}\cdot{O})=\vec{V}\cdot\vec{r}$, and thus if we set $\vec{V}||\vec{r}$, we immediately see that $\sum_{i=1}^n\langle O_i\rangle^2=\vec{r}^2\leq 1$. 
\begin{flushright}
$\blacksquare$
\end{flushright}

\begin{fact}

Given hermitian operators $O_1,...,O_n$ with eigenvalues $\pm 1$, which all commute with each other, the maximum of $\sum_{i=1}^n\langle O_i\rangle^2$ is $n$.
\end{fact}
Naturally, this maximum is obtained by taking the average with respect to the common eigenstates of $O_{i}$'s. We will also use the following fact:

\begin{fact}

Linear function obtains its maximum in one or more extreme points of its domain.
\end{fact}
The function we will try to maximize is, indeed, a linear function (in fact, basically a sum) of variables $a_i=\langle O'(s_i)\rangle^2$. The bounds of the domain will also have a linear form, being due either to the spectral properties of the operators, $0\leq a_i\leq 1$, or to the anticommutation relations, say, $a_1+a_3\leq 1$.

\begin{lemma}

Mixing states cannot make the value of the left-hand side (LHS) of Eq. (\ref{bellcrit}) larger.
\end{lemma}

{\em Proof:} $Q$ is essentially a quadratic, positive semi-definite form, which is familiar from Ref. \cite{EXPFRIEN},
\begin{equation}
Q=(T,T)=\sum_{i,j,...,k=0}^3T_{ij...k}^2G_{ij...k}.
\end{equation}
The ``metric'' tensor $G_{ij...k}$ has only entries 0 or 1, and $T_{ij...k}$ are elements of the correlation tensor (see Ref. \cite{ZB}). Let us now take a mixture $\rho=p\rho_1+(1-p)\rho_2$, which corresponds to $T=pT(1)+(1-p)T(2)$. Assume that $(T(1),T(1))\geq(T(2),T(2))$. Then because of the positivity of $(T(1)-T(2),T(1)-T(2))$ we have
\begin{eqnarray}
(T(1),T(1))&\geq&\frac{1}2((T(1),T(1))+(T(2),T(2)))\nonumber\\
&\geq&(T(1),T(2)).
\end{eqnarray}
Note that $Q$ would be a mixture of $(T(1),T(1))$, $(T(2),T(2))$, and $(T(1),T(2))$ with respective weights $p^2,(1-p)^2,$ and $2p(1-p)$. Hence $Q$ is always smaller than $(T(1),T(1))$ unless $p=1$.
\begin{flushright}
$\blacksquare$
\end{flushright}

Finally, we define the cut-anticommutativity and cut-commutativity.
\begin{definition}

Given two operators, $C_1=A_1B_1$ and $C_2=A_2B_2$, which are products with respect to cut $A|B$, we say that they $A|B$-anticommmute if they anticommute on part $A$, $\{A_1,A_2\}=0$ or on part $B$, $\{B_1,B_2\}=0$.
\end{definition}

\begin{definition}

Given two operators, $C_1=A_1B_1$ and $C_2=A_2B_2$, which are products with respect to cut $A|B$, we say that they $A|B$-commute if they commute on part $A$, $[A_1,A_2]=0$ and on part $B$, $[B_1,B_2]=0$.
\end{definition}

For example, operators $xx$ and $yy$ commute, but not with respect to the cut. Operators $x1x1$ and $xy1z$ commute with respect to all cuts.

With these definitions, we now present the last two of the Lemmas.
\begin{lemma}

Consider a set of hermitian operators with eigenvalues $\pm 1$, $\{A_i B_i\}_{i=1}^M$ such that for all $i\neq j$ $A_iB_i$ $(A|B)$-anticommutes with $A_jB_j$. Additionally, every pair $A_i,A_j$ and $B_i,B_j$ either commutes or anticommutes. Then for all states separable with respect to cut $A|B$ we shall have $\sum_{i=1}^M\langle A_i B_i\rangle^2\leq 1$.
\end{lemma}
{\em Proof:} By Lemma 2 we focus only on product states. Note that if a unit matrix appears as one of the operators $A_i$ it cannot appear as one of the $B_i$'s. Without loss of generality, we shall call the side, in which the unit operator does not appear, $B$, thus all operators $B_i$ anticommute. Let us group the operators in terms of commutativity on the side $A$, so that all operators $A_i^{(k)}$ in the same group $k$ commute with each other, i.e., $[A_i^{(k)},A_j^{(k)}]=0$. Each operator of $B_i$ shall also be labeled with $k$, depending on the group its partner $A_i$ is in. We also have
\begin{eqnarray}
\{A_i^{(k)},A_j^{(l)}\} &=& 0\quad (k\neq l),\nonumber\\
\{B_i^{(k)},B_j^{(k)}\}&=&0.
\end{eqnarray}
Now, because of the anticommutativity of $B_i^{(k)}$'s within each group we have
\begin{equation}
\sum_i\langle A_i^{(k)}\rangle^2\langle B_i^{(k)}\rangle^2\leq \langle \bar{A}_{k}\rangle^2,
\end{equation}
where $\langle \bar{A}_k\rangle^2=\max_{i}\langle A_i^{(k)}\rangle^2$, i. e., we choose $\bar{A}_k$ from the set of $A_i^{(k)}$'s. Since operators $\bar{A}_{k}$ anticommute with each other, we obtain $\sum_{k}\langle \bar{A}_k\rangle^2\leq 1$ and hence the conclusion of the lemma is proven.
\begin{flushright}
 $\blacksquare$
\end{flushright}

\begin{lemma}
A product state with respect to cut $(A|B)$ can be a common eigenstate of hermitian operators $A_1B_1,...,A_nB_n$ (being products with respect to this cut) with eigenvalues $\pm 1$ only if all these operators $(A|B)$-commute with each other. In such a case $\sum_{i=1}^n\langle A_i B_i\rangle^2\leq n$, with the bound saturated for the common eigenstates.
\end{lemma}
Under the restriction that all operators either commute or anticommute, the proof follows straightforwardly from lemma 3.

We define a set of operators $\sigma$ containing only $N$-fold tensor products of Pauli matrices. Two such products either commute or anticommute.  Let our criterion take the form
\begin{equation}
Q=\sum_{O\in\sigma}(\text{tr}\rho O)^2<B_{SC}.
\end{equation}
$B_{SC}$ is a bound dependent on the separability class (SC). If this bound is beaten, the state $\rho$ cannot belong to this class.
The facts and lemmas above validate the following procedure to establish the maxima of $Q$ obtained by states belonging to various SCs. We will assign values either 1 or 0 to terms of $Q$, keeping in mind the restriction imposed by the (cut-)anticommutation relations. The maximum of $Q$ for a given cut will be given by the independence number of the cut-anticommutativity graph for operators belonging to $\sigma$. The independence number is defined as the maximal possible number of vertices that are disconnected with each other. It is equal to the clique number of the cut-commutativity graph with the same partition. The clique number is the number of vertices in the largest subgraph (\textit{clique}), all of whose vertices are mutually connected. By definition, commutativity graphs are a complement of anticommutativity graphs with the same cut.

We proceed with an example of a necessary condition for the full separability for three qubits studied in \cite{WW,WZ,ZB},
\begin{equation}
\sigma=\{xxx,yxx,xyx,yyx,xxy,yxy,xyy,yyy\}.
\end{equation}
The anticommutation graph of $\sigma$ takes the form of a cube with its all four diagonals. Hence the independence number of the graph is 4 --- the vertices of a tetrahedron inscribed into the cube are not directly connected with each other. If we take states separable with respect to one of the qubits, new edges appear on the cut-anticommutativity graphs --- diagonals of a pair of parallel faces. This reduces the independence number to 2. Finally, if we insist on the full separability of the state, all face diagonals appear as edges, giving the maximum for fully separable states equal to 1. This analysis for more particles would allow us to easily reproduce the result of \cite{YuChen}.

Another example we present here is the operator set that was introduced in Ref. \cite{WNZ} to study the Bell inequalities with lower order correlations. Consider
\begin{eqnarray}
\label{sigma}
\sigma=CP(1xxxz)\cup CP(1zxxz)\cup CP(1zxzz),
\end{eqnarray}
where 1 denotes the local unit matrix and $CP$ stands for a set of cyclic permutations of the argument. This symmetry of $\sigma$ will allow us to segregate all bipartite states into only three classes.

First we are going to show
\begin{proposition}

For fully separable states, the maximum value of $Q$ is 1.
\end{proposition}
The construction of $\sigma$ is presented in details in Ref. \cite{WNZ}. The important feature of th elements of $\sigma$ is that for every pair of them there is such a position, at which one of the elements has $x$, while the other has $z$. Hence they all $(A|B|C|D|E)$-anticommute; a theorem similar to Lemma 2 applies here.
By Lemma 2, we focus only on fully separable product states. In order to evaluate $Q$, let us write
\begin{eqnarray}
\langle z1111\rangle^2\leq a,&\quad&\langle x1111\rangle^2\leq (1-a),\nonumber\\
\langle 1z111\rangle^2\leq b,&\quad&\langle 1x111\rangle^2\leq (1-b),\nonumber\\
\langle 11z11\rangle^2\leq c,&\quad&\langle 11x11\rangle^2\leq (1-c),\nonumber\\
\langle 111z1\rangle^2\leq d,&\quad&\langle 111x1\rangle^2\leq (1-d),\nonumber\\
\langle 1111z\rangle^2\leq e,&\quad&\langle 1111x\rangle^2\leq (1-e),
\end{eqnarray}
with $0\leq a,b,c,d,e\leq 1$. Then, $Q$ is bounded by
\begin{widetext}
\begin{eqnarray}
Q&\leq& abc(1-d)+bcd(1-e)+(1-a)cde+a(1-b)de+ab(1-c)e\nonumber\\
&+&(1-a)bc(1-d)+ab(1-c)(1-e)+a(1-b)(1-d)e+(1-a)(1-c)de+(1-b)cd(1-e)\nonumber\\
&+&(1-a)b(1-c)(1-d)+a(1-b)(1-c)(1-e)+(1-a)(1-b)(1-d)e\nonumber\\
&+&(1-a)(1-c)d(1-e)+(1-b)c(1-d)(1-e)\nonumber\\
&\leq& abc(d-1)+bcd(1-e)+(1-a)cde+a(1-b)de+ab(1-c)e\nonumber\\
&+&(1-a)bc(1-d)+ab(1-c)(1-e)+a(1-b)(1-d)e+(1-a)(1-c)de+(1-b)cd(1-e)\nonumber\\
&+&(1-a)b(1-c)(1-d)+a(1-b)(1-c)(1-e)+(1-a)(1-b)(1-d)e\nonumber\\
&+&(1-a)(1-c)d(1-e)+(1-b)c(1-d)(1-e)\nonumber\\
&+&abcde+(1-a)(1-b)(1-c)(1-d)(1-e)\nonumber\\
&=&1.
\end{eqnarray}
\end{widetext}
Now, let us proceed to more generic cases, demonstrating bounds on $Q$ for various states that may have different entanglement structures. We shall label the qubits with $A,B,C,D,E$. Figures 1, 2, 3 present graphs of cut-commutativity relations between operators from $\sigma$ with respect to cuts $(A|BCDE)$, $(AB|CDE)$, and $(AC|BDE)$, respectively. Due to the cyclic permutation symmetry of $\sigma$, any two-partite cut will be equivalent to one of these cases.
Only those operators that are represented by vertices directly connected can be simultaneously assigned values $\pm 1$. Thus the maximal value of $Q$ is given by the size of the largest clique in the graph. One can easily verify that in Fig. 1 there are cliques of three vertices. In the other graphs in Figs. 2 and 3 all cliques are simply edges. Hence $Q\leq 3$ for all bipartite separable states.

%\begin{centering}
\begin{figure}[t]
\includegraphics[width=5cm]{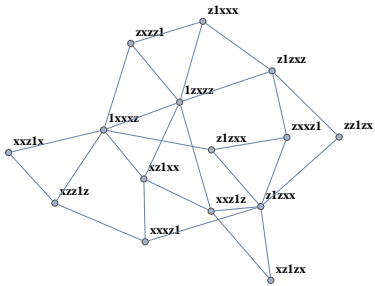}
\caption{The $(A|BCDE)$-commutation diagram for operators from from set (\ref{sigma}). Due to the cyclic permutation symmetry of $\sigma$, analogous graphs apply to all other (4|1)-cuts.}
\end{figure}
%\begin{centering}
\begin{figure}[t]
\includegraphics[width=5cm]{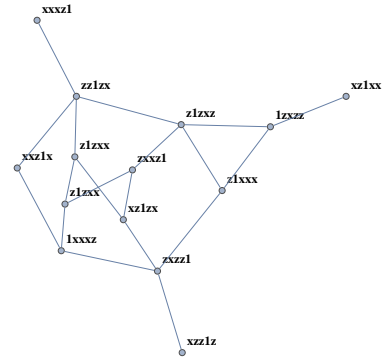}
\caption{The $(AB|CDE)$-commutation diagram for operators from set (\ref{sigma}). Due to the cyclic permutation symmetry of $\sigma$, analogous graphs apply to all other (3|2)-cuts with two neighboring qubits at one side.}
\end{figure}
\begin{figure}[t]
\includegraphics[width=5cm]{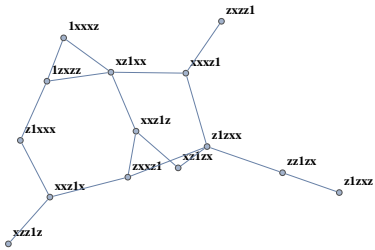}
\caption{The $(AC|BDE)$-commutation diagram for operators from set (\ref{sigma}). Due to the cyclic permutation symmetry of $\sigma$, similar graphs apply to all other (3|2)-cuts with two non-neighboring qubits at one side.}
\end{figure}

On the other hand, Fig. 4 shows a graph of commutativity relations between the operators in $\sigma$. The biggest clique found in this graph is of 5 elements, namely $zxxz1$, $xxz1z$, $xz1zx$, $z1zxx$ and $1zxxz$. Hence $Q$ can attain 5.

\begin{figure}[t]
\includegraphics[width=5cm]{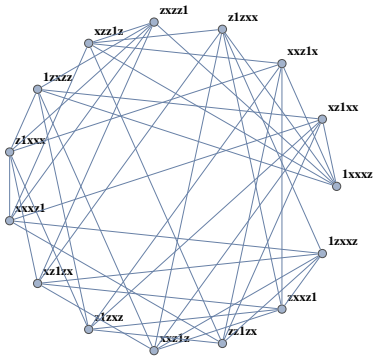}
\caption{The commutation diagram for operators from set (\ref{sigma}).}
\end{figure}

We have thus shown the following.
\begin{proposition}

For $\sigma$ given by Eq. (\ref{sigma}), in a given state of five qubits $\rho$, $Q>1$ implies the presence of entanglement in $\rho$. In addition $3<Q\leq 5$ requires genuine five-partite entanglement in $\rho$.
\end{proposition}

Note that the state $\rho$ given under Eq. (9) in Ref. \cite{WNZ} violates a Bell inequality based on $\sigma$ of Eq. (\ref{sigma}) by a factor of 1.806. By the Cauchy-Schwartz inequality, this implies $Q>3$ (see Ref. \cite{WNZ}). Thus $\rho$ is genuinely five-partite-entangled despite the absence of 5-partite correlations.

In conclusion, we have shown a useful toolbox for formulating (multipartite) entanglement criteria. These criteria take the form of an inequality, with a separability class-dependent upper bound on one side, and an incomplete sum of squared correlation tensor elements on the other. Importantly, our method eliminates the necessity of finding the optimal state maximizing the expression as in Refs. \cite{EXPFRIEN,EXPFRIEN1}.

We have also shown an example with a set of operators that was studied in Ref. \cite{WNZ} in the context of the Bell inequalities with lower-order correlations. Interestingly, we are able to test $N$-partite entanglement without using $N$-partite correlations. It is important to note that non-classical lower-order correlations are characteristic of W-type entangled states, whereas they do not appear in GHZ-like states. Hence we can additionally argue that the entanglement criterion given in example 2 does not detect 5-partite GHZ entanglement.

Lastly, let us stress that we have presented a family of multipartite entanglement criteria, rather than a single one of them, given by all possible sets of observables $\sigma$'s. Those $\sigma$'s with only three observables at each side  (say $x$, $y$, and $1$) can be obtained by considering the problem of matching vertices of $N$-cube graph into groups of 1, 2, 4... It is known that the number of such matching grows superexponentially with respect to $N$ \cite{PROPP}. Therefore, our toolbox is highly efficient in producing a large number of multipartite entanglement criteria in a straightforward and transparent manner.

\begin{acknowledgments} 
The project was part of the EU project QUASAR. MW is supported by the Foundation for Polish Science (HOMING PLUS Program). KM is grateful for the support of the JSPS Kakenhi (C) no. 22540405.
\end{acknowledgments}

\end{document}